\RequirePackage{lineno} 
\documentclass[12pt,twoside]{article}
\usepackage{calc}
\usepackage{graphicx}
\usepackage{here}
\usepackage{color}
\usepackage{amsmath}
\usepackage{longtable}
\usepackage{graphics}
\usepackage{subfigure}
\usepackage{setspace}
\usepackage{units}
\usepackage[margin=1in]{geometry}
\usepackage{dcolumn}
\usepackage{bm}
\graphicspath{{figures/}}

\def\g2{{$(g-2)$} }
\def\be{\begin{equation}}
\def\ee{\end{equation}}
\def\ben{\begin{enumerate}}
\def\een{\end{enumerate}}
\def\bi{\begin{itemize}}
\def\ei{\end{itemize}}
\def\bs{\begin{slide}}
\def\es{\end{slide}}
\def\bea{\begin{eqnarray} }
\def\eea{\end{eqnarray} }
\def\bc{\begin{center} }
\def\ec{\end{center} }

\newcommand{\mev}{\mbox{MeV}}

\begin{document}

\title{The Muon $(g-2)$ Theory Value: Present and Future}

\author{Thomas Blum$^1$,  Achim Denig$^{2}$,
 Ivan  Logashenko$^3$,  Eduardo de Rafael$^4$,\\ 
B. Lee Roberts$^5$, Thomas Teubner$^6$, Graziano Venanzoni$^7$ \\
\\
$^1$Department of Physics, University of Connecticut,\\
 and the 
RIKEN BNL Research Center, USA\\
$^2$Johannes Gutenberg Universit\"at, Institut f\"ur Kernphysik \\
and PRISMA,
Mainz, Germany\\ 
$^3$Budker Institute of Nuclear Physics, Novosibirsk, Russia\\
$^4$Centre de Physique Th\'eorique, CNRS-Luminy, Case 907, \\
F-13288 Marseille Cedex 9, France\\
$^5$Department of Physics, Boston University, 
Boston, MA, USA\\
$^6$Department of Mathematical Sciences, University of Liverpool, Liverpool,
UK\\ 
$^7$Laboratori Nazionali di Frascati dell'INFN, Frascati, Italy\\}
 
\vskip0.5in

\maketitle
                     
\singlespacing                                                           

\begin{abstract}

This White Paper briefly reviews the present status of the muon \g2
Standard-Model prediction. This value results in a 3 -- 4 standard-deviation
difference with the experimental result from Brookhaven E821. 
The present experimental uncertainty is $\pm 63 \times 10^{-11}$ (0.54~ppm), and
the Standard-Model uncertainty is $\simeq \pm 49 \times 10^{-11}$.
Fermilab experiment E989 has the goal to reduce the 
experimental error to $\pm 16 \times 10^{-11}$. 
Improvements in the Standard-Model value, which 
should be achieved between now and when the
first results from Fermilab E989 could be available,
should lead to a Standard-Model 
uncertainty of $\sim \,\pm 35 \times 10^{-11}$.  These improvements would
halve the uncertainty on the difference between experiment and theory, and
should 
clarify whether the current difference points toward 
 New Physics, or to a statistical fluctuation.
  At present,
 the \g2 result is arguably the most compelling indicator of
physics beyond the Standard Model and, at the very least, it
represents a major constraint for speculative new theories such as
supersymmetry, dark gauge bosons or extra dimensions. 
\end{abstract}

\newpage

\section{Introduction}

%
%
%

The Standard-Model (SM) value of the muon anomaly can be calculated with 
sub-parts-per-million precision.  The comparison between the measured and
the SM prediction provides a test of the completeness of the
Standard Model.  At present, there appears to be a three- to four-standard
deviation between these two values, which has motivated extensive theoretical
and experimental work on the hadronic contributions to the muon anomaly.

\begin{figure}[h!]
\begin{center}
  \includegraphics[width=0.7\textwidth,angle=0]{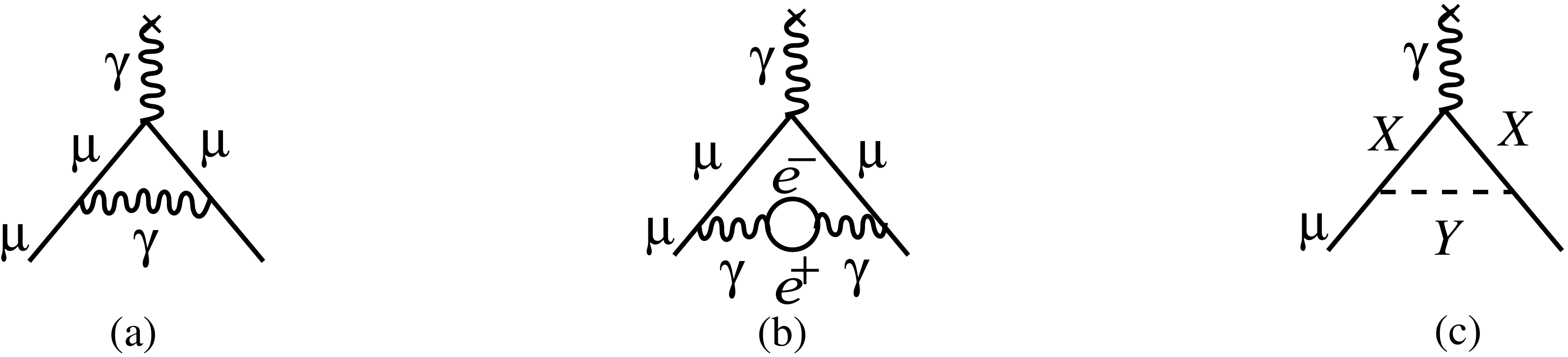}
  \caption{The Feynman graphs for: (a) The lowest-order (Schwinger)
contribution to the lepton anomaly ; (b)  The vacuum polarization
contribution, which is one of five fourth-order,
$( \alpha/\pi)^2$, terms; (c) The schematic contribution of new particles
$X$ and $Y$ that couple to the muon. 
  \label{fg:schwinger}}
\end{center}
\end{figure}

A lepton ($\ell = e,\,\mu,\,\tau$) has a magnetic moment which is along its
spin, given by the relationship
 \be \vec \mu_{\ell} =g_{\ell}
\frac{Qe}{2m_{\ell}}\vec{s}\,,\qquad \underbrace{g_{\ell} =2}_{\mbox{\rm
\small Dirac}}(1+a_\ell), \qquad a_\ell = \frac{g_\ell -2}{2} 
\ee 
where $Q = \pm 1$, $e>0 $ and $m_\ell$ is the lepton mass.  
Dirac theory predicts
that $g \equiv 2$, but experimentally, it is known to be greater than 2.
 The small number $a$, the anomaly, arises from quantum fluctuations,
with the largest contribution coming from the mass-independent
 single-loop diagram in
Fig.~\ref{fg:schwinger}(a). With his famous calculation 
that obtained  $a = (\alpha/2 \pi) = 0 .00116\cdots$,
Schwinger~\cite{Schwinger:1948} started an ``industry'',
which required  Aoyama, Hayakawa, Kinoshita and Nio
to calculate more than 12,000 diagrams to evaluate the tenth-order
(five loop) contribution~\cite{Aoyama1:2012}.   

The interaction shown in Fig.~\ref{fg:schwinger}(a) is a
chiral-changing, flavor-conserving process, which gives it a special
sensitivity to possible new physics~\cite{Miller:2012,Stoeckinger2010}. 
Of course heavier particles can also contribute, as indicated by the
 diagram in Fig.~\ref{fg:schwinger}(c).  For example,
  $X = W^\pm$ and $Y = \nu_\mu$, along with
$X = \mu$ and $Y = Z^0$, are the lowest-order weak contributions.  
 In the
Standard-Model, $a_\mu$ gets measureable
 contributions from QED, the strong interaction, and
from the electroweak interaction,
\be
a^{SM} = a^{QED} + a^{Had} + a^{Weak} .
\ee

In this document we present the latest evaluations of the SM
value of $a_\mu$, and then discuss expected improvements
that will become available over the next five to seven years. 
The uncertainty in this evaluation is dominated by the
contribution of virtual hadrons in loops.  A worldwide effort is under way to
improve on these hadronic contributions. By the time that 
the Fermilab muon $(g-2)$ experiment, E989, reports a result later in this
decade, the uncertainty should be significantly reduced. We emphasize that
the existence of E821 at Brookhaven motivated significant work over the past
thirty years that permitted more than an order of magnitude improvement in the
knowledge of  the hadronic contribution. Motivated by
 Fermilab E989 this work continues, and another factor of two
improvement could be possible. 

Both the electron~\cite{Hanneke08}
 and muon~\cite{Bennett:2006I} anomalies have been measured very precisely:
\bea
a_{e}^{exp}   &=& 1\,159\,652\,180.73 \,(28)\times 10^{-12} \ \ {\pm 0.24\,
  \rm{ppb}}  \\ 
a_{\mu}^{exp} &=& 1\,165\,920\,89\, (63) \times 10^{-11} \ \ {\pm 0.54\,
  \rm{ppm}}  
\eea
While the electron anomaly has been measured to $\simeq 0.3$~ppb (parts per
billion)~\cite{Hanneke08}, it is significantly less
sensitive to heavier physics,
because the relative
contribution  of heavier virtual particles to the muon anomaly goes
as $(m_\mu/m_e)^2 \simeq 43000$.  Thus the lowest-order hadronic contribution
to $a_e$ is~\cite{Davier-LM}:
$ a_e^{\rm had,LO} = (1.875\pm0.017)~10^{-12}$, 1.5~ppb
of $a_e$.  For the muon the hadronic contribution is $\simeq 60$~ppm 
(parts per million).  So with much less precision, when
compared with the electron, the measured muon anomaly is sensitive to mass
scales in the several hundred GeV region.  This not only includes
the  contribution of the $W$ and $Z$ bosons, but perhaps
contributions from new, as yet undiscovered, particles such as the
supersymmetric partners of the electroweak gauge bosons (see
Fig.~\ref{fg:schwinger}(c)).

\section{Summary of the Standard-Model Value of $a_\mu$}


\subsection{QED Contribution}

The QED contribution to $a_\mu$ is  well
understood.  Recently the four-loop QED contribution has been updated
 and the full five-loop contribution has been calculated~\cite{Aoyama1:2012}.
The present QED value is
\be
a_{\mu}^{\rm QED} =  116~584~718.951~(0.009)( 0.019 )(0.007)(.077)
 \times 10^{-11}
\ee
where the  uncertainties are from the lepton mass ratios, the eight-order
term, the tenth-order term, and the value of
  $\alpha$  taken from the $^{87}$Rb  atom
 $\alpha^{-1}(Rb) = 137.035\,999\,049(90)$ [0.66 ppb].~\cite{Bouchendira:2011}.

\subsection{Weak contributions}

\begin{figure}[h!]
\begin{center}
  \includegraphics[width=0.8\textwidth]{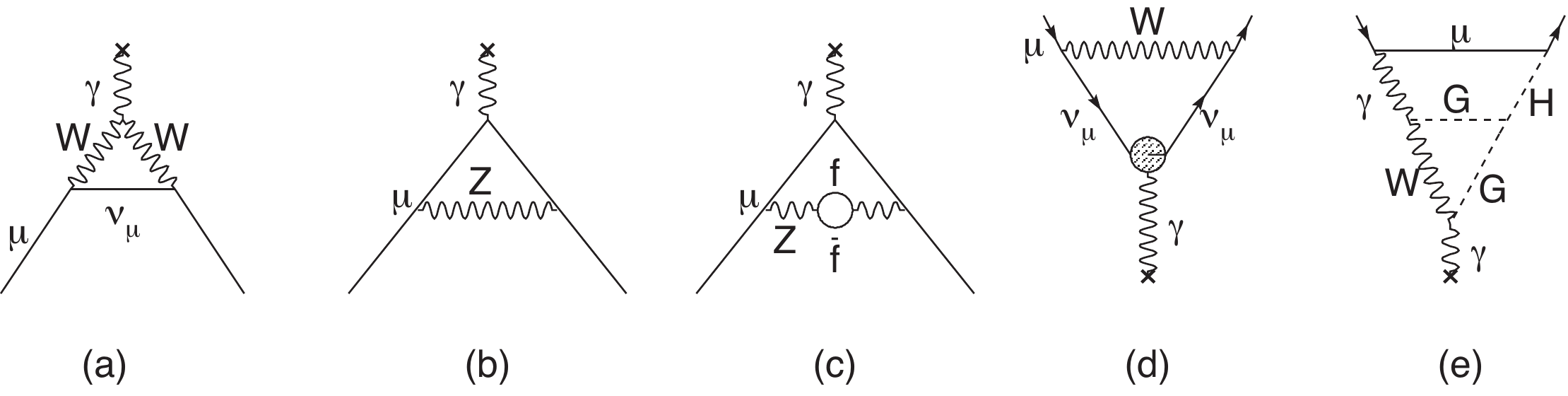}
\end{center}
  \caption{Weak contributions to the muon anomalous magnetic moment.
Single-loop contributions from
(a) virtual $W$ and (b) virtual $Z$ gauge bosons.
These two contributions enter with
opposite sign, and there is a partial cancellation.
The two-loop contributions fall into three categories:
(c) fermionic loops which involve the coupling of the gauge bosons to quarks,
(d) bosonic loops which appear as corrections to the one-loop diagrams, and
(e) a new class of diagrams involving the Higgs
where {\sl G} is the longitudinal component of the
gauge bosons.  See Ref. \cite{Miller:07I} for details.
The $\times$
indicates the  photon from the magnetic field.}
\label{fg:weak}
\end{figure}

The electroweak contribution (shown in Fig.~\ref{fg:weak})
is now calculated through two 
loops~\cite{CKM96,PPdR95,CKM95,CMV03,CM-LM,Gnendiger:2013}.
The one loop result
\bea
\label{EW1}
a_{\mu}^{\mbox{\rm\tiny
 EW(1)}} &\!\! =\!\! &
\frac{G_{\mbox\tiny F}}{\sqrt{2}}\frac{m_{\mu}^2}{8\pi^2}
\left\{\underbrace{\frac{10}{3}}_{W} +
\underbrace{\frac{1}{3}(1\!-\!4\sin^{2}\theta_{W})^2-\frac{5}{3}}_{Z}  \right.
\nonumber
\\
&  + & \!\!\left. {\mathcal O} \!\left(\!
\frac{m_{\mu}^2}{M_{Z}^2}\log\frac{M_{Z}^2}{m_{\mu}^2}\!\right)
\!+\!\frac{m_{\mu}^2}{M_{H}^2}
\int_{0}^{1}\!\! dx
\frac{2x^2
(2-x)}{1-x+{\frac{m_{\mu}^2}{M_{H}^2}}x^2}
\!\right\} \nonumber  \\
 & = & 194.8\times 10^{-11}\,,
\eea 
was calculated by five separate groups~\cite{weak} shortly after the
Glashow-Salam-Weinberg theory was shown by 't Hooft to be
renormalizable. Due to the small Yukawa coupling of the Higgs boson to the 
muon, only the $W$ and $Z$ bosons contribute at a measurable level
in the lowest-order electroweak term.

The two-loop electroweak contribution (see Figs.~\ref{fg:weak}(c-e)),
 which is negative~\cite{CKM95,PPdR95,CKM96,CMV03},
 has been re-evaluated using the
 LHC value of the Higgs mass~\cite{Gnendiger:2013}.
The total electroweak contribution is
\be
a_{\mu}^{\rm EW} = (153.6\pm 1.0) \times 10^{-11}
\label{eq:ew}
\ee
where the  error comes from hadronic effects in the second-order
electroweak diagrams with quark triangle loops, along with unknown three-loop
contributions\cite{CMV03,KPPdeR02,Vain03,KPPdeR04}.  The leading
logs for the next-order term have been shown to be
small~\cite{CMV03,Gnendiger:2013}. The weak contribution is about 1.3~ppm of the
anomaly, so the experimental  uncertainty on $a_\mu$ of $\pm
0.54$~ppm now probes the weak scale of the Standard Model.

\subsubsection{Hadronic contribution}

\begin{figure}[h!]
\begin{center}
  \includegraphics[width=32pc]{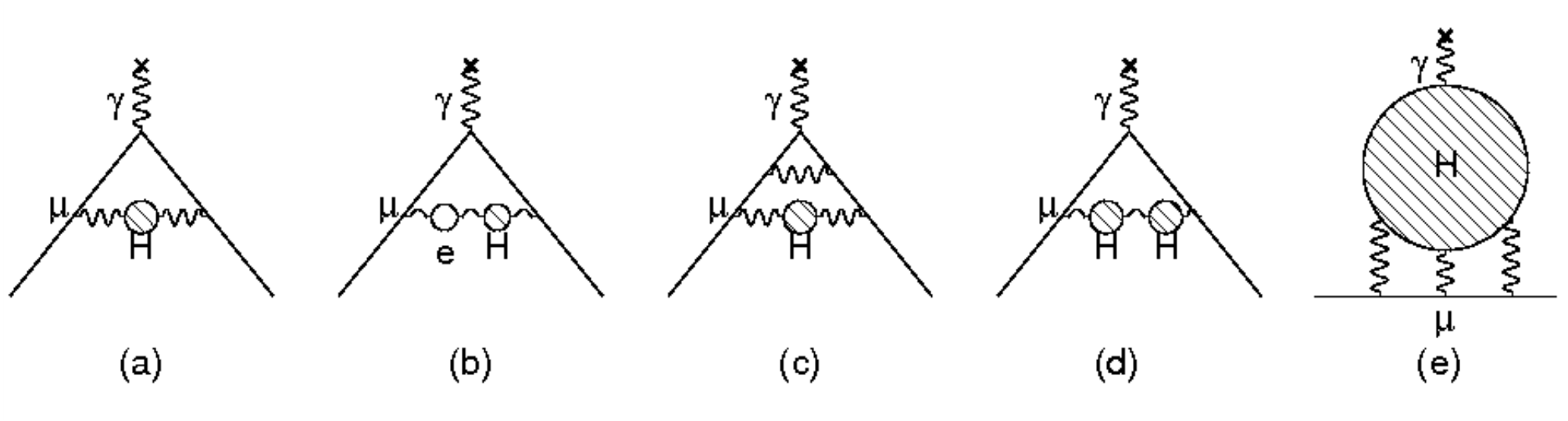}
\end{center}
  \caption{The hadronic contribution to the muon anomaly, where the
dominant contribution comes from the lowest-order diagram (a).  The
 hadronic light-by-light contribution is shown in (e).}
\label{fg:had}
\end{figure}

The hadronic contribution to $a_{\mu}$ is about 60 ppm of the total
value.
 The lowest-order diagram shown in Fig.~\ref{fg:had}(a)
dominates this contribution and its error, but the hadronic light-by-light
contribution Fig.~\ref{fg:had}(e) is also important.  We discuss both of
these contributions below.

\begin{figure}[h!]
\begin{center}
  \includegraphics[width=0.54\textwidth]{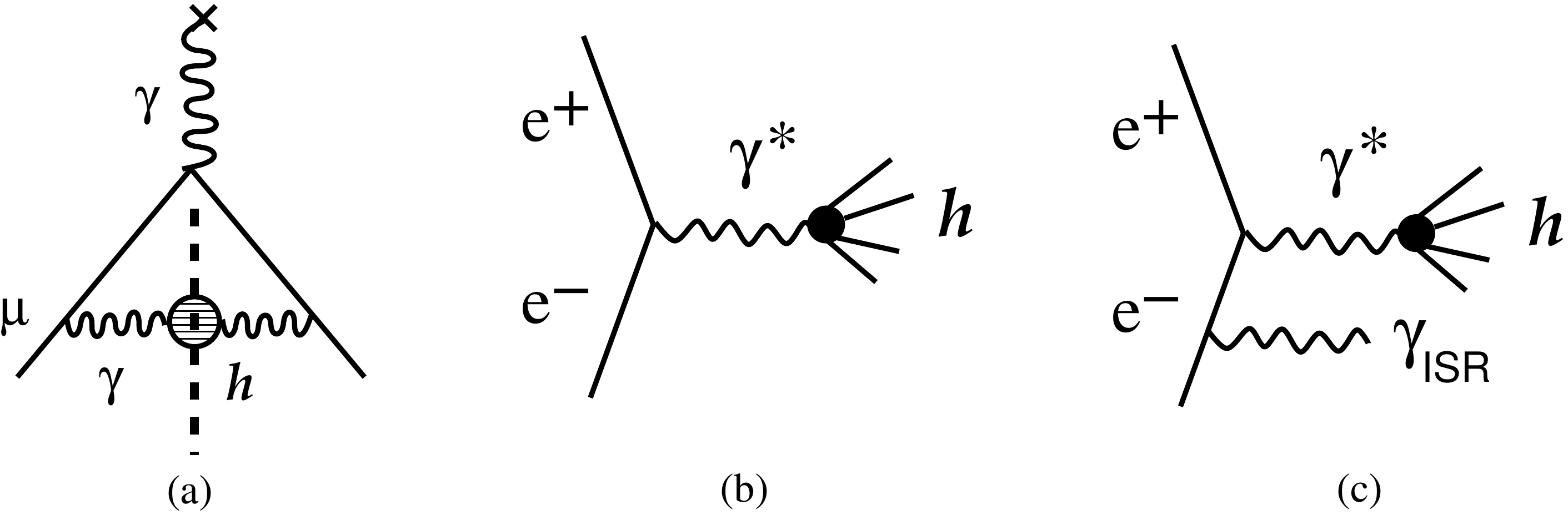}
\end{center}
  \caption{(a) The ``cut'' hadronic vacuum polarization diagram; (b) The
$e^+ e^-$ annihilation into hadrons; (c) Initial state radiation
accompanied by the production of hadrons.}
\label{fg:hadpro}
\end{figure}

\begin{figure}[h!]
\vskip0.2in
\begin{center}
  \includegraphics[width=0.5\textwidth]{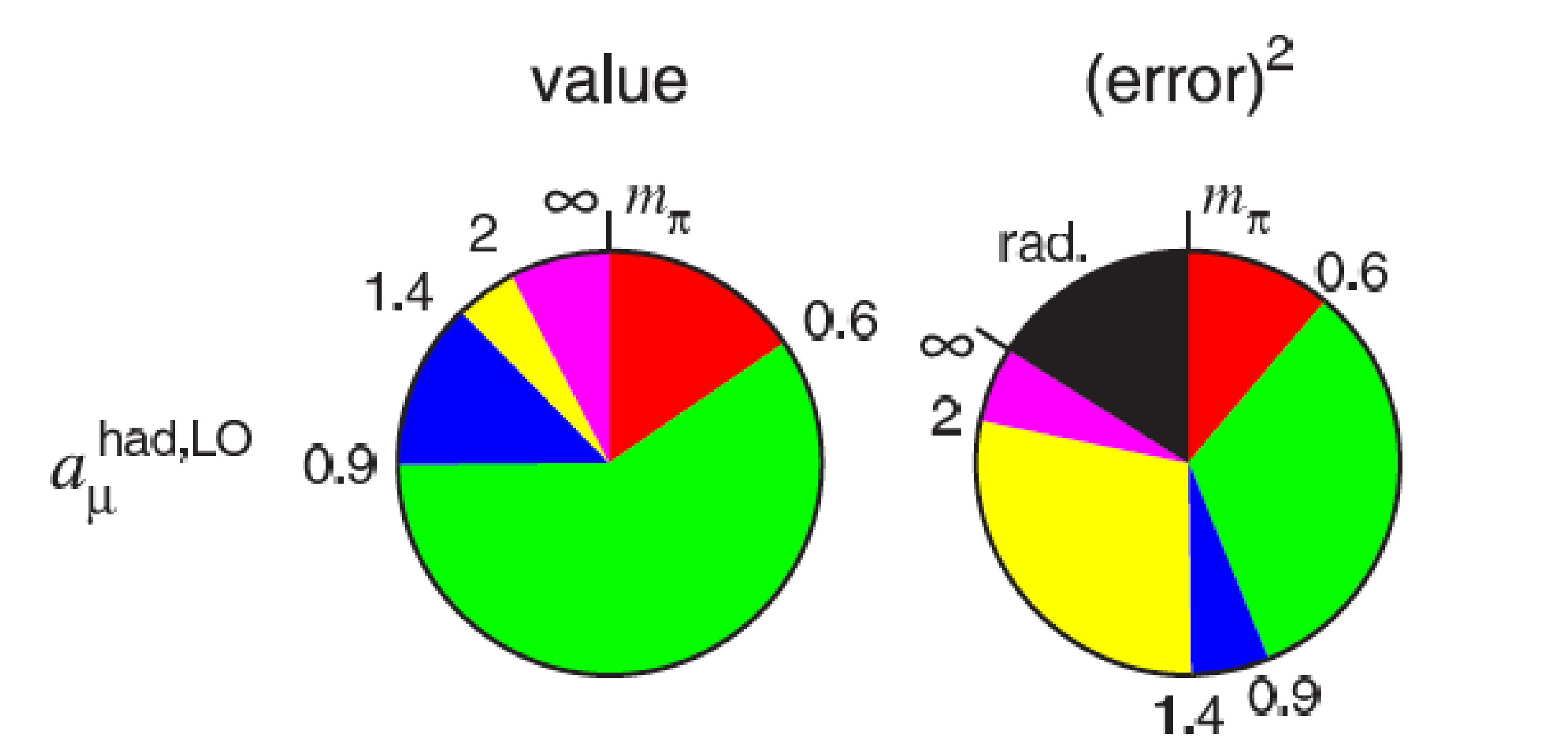}
\end{center}
  \caption{Contributions to the dispersion integral for different
energy regions, and to the associated error 
(squared) 
on the dispersion integral in that energy region. Taken from
 Hagiwara et al.~\cite{Hagiwara:2011}.}
\label{fg:had-cont}
\end{figure}

The energy scale for the virtual hadrons is of order $ m_{\mu} c^2$,
well below the perturbative region of QCD. However it can be 
calculated from the
dispersion relation shown pictorially in Fig.~\ref{fg:hadpro},
\begin{equation}
a_{\mu}^{\rm had;LO}=\left({\alpha m_{\mu}\over 3\pi}\right)^2
\int^{\infty} _{m_{\pi}^2} {ds \over s^2}K(s)R(s), \quad
{\rm where} \quad
R\equiv{ {\sigma_{\rm tot}(e^+e^-\to{\rm hadrons})} \over
\sigma_{\rm }(e^+e^-\to\mu^+\mu^-)}\, ,
\label{eq:dispersion}
\end{equation}
 using
the measured cross sections for  $e^+ e^- \rightarrow {\rm hadrons}$
as input, where $K(s)$ is a kinematic factor ranging from { 0.4} at
$s= m_\pi^2$ to $0$ at $s = \infty$ (see Ref.~\cite{Miller:07I}).
 This dispersion relation relates the bare cross
section for $e^+e^- $ annihilation into hadrons
to the hadronic vacuum polarization contribution
to $a_{\mu}$.
Because the integrand
contains a factor of $s^{-2}$, the values
of  $R(s)$ at low energies (the $\rho$ resonance) dominate the determination of
$a_{\mu}^{\rm had;LO}$, however at the level of precision needed, the
 data up to 2~GeV are very important.
This is shown in Fig.~\ref{fg:had-cont}, where the
left-hand chart gives the relative contribution to the integral for the
different energy regions, and the right-hand gives the contribution to the
error squared on the integral.
The contribution is dominated by the two-pion final
state, but other low-energy multi-hadron cross sections are also important.
 These data for $e^+e^-$ annihilation to hadrons are also important as
input into the determination of $\alpha_{QED}(M_Z)$ and other electroweak
precision measurements.

Two recent analyses~\cite{Davier11,Hagiwara:2011}
 using the $e^+e^- \to hadrons$ data obtained:
 \bea
a_{\mu}^{\rm had;LO}&=&\left(6~923\pm 42\right)\times
10^{-11}\,, \label{eq:hvp1-pub}\\
a_{\mu}^{\rm had;LO}&=&\left(6~949\pm 43\right)\times
10^{-11}\,, 
\label{eq:hvp2-pub}
\eea
respectively.
Important earlier global analyses include
those of Hagiwara et al.~\cite{HMNT07}, Davier, et al.,~\cite{Davier07},
Jegerlehner and Nyffler~\cite{Jegerlehner:2009ry}.

In the past, hadronic $\tau$ spectral functions and CVC, together with
isospin breaking corrections have been used to calculate the hadronic
contribution~\cite{Alemany:1998,Davier11}.
While the original predictions showed a discrepancy between
$e^+e^-$ and $\tau$ based evaluations, it
has been shown that after $\gamma$-$\rho$ mixing is taken into account, the
two are compatible~\cite{Jegerlehner:2011}. Recent evaluations based on a
combined $e^+e^-$ and $\tau$ data fit using the Hidden Local Symmetry (HLS)
model have come to similar conclusions and result in values for $a_\mu^{\rm
  HVP}$ that are smaller than the direct evaluation without the HLS 
fit~\cite{Benayoun:2012a,Benayoun:2012b}.

The most recent evaluation of the next-to-leading 
order hadronic contribution shown in
 Fig.~\ref{fg:had}(b-d), which can also be determined
from a dispersion relation, is~\cite{Hagiwara:2011} 
\be
\label{eq:exphvpnlo} 
a_{\mu}^{\rm had;NLO}=(-98.4\pm 0.6_{\exp}\pm0.4_{\rm rad}\ )\times 10^{-11}\,.
\ee

\subsubsection{Hadronic light-by-light contribution}


The hadronic light-by-light contribution (HLbL)
cannot at present be determined from data, but rather must be
calculated using hadronic models that correctly reproduce
properties of QCD. This contribution is shown below in
Fig.~\ref{fg:HLBL}(a). It is dominated by the long-distance contribution
shown in Fig.~~\ref{fg:HLBL}(b). In fact, in the so called  chiral limit
where  the mass gap between the pseudoscalars ( Goldstone-like) particles and
the other hadronic particles (the $\rho$ being the lowest vector
state in Nature) is
considered to be large, and to leading order in the $1/N_c$--expansion ($N_c$
the number of colors), this contribution has been calculated
analytically~\cite{KNPdeR02} and provides a long-distance constraint to model
calculations. There is also a short-distance constraint from the operator
product expansion (OPE) of two  electromagnetic currents which, in specific
kinematic conditions, relates the light-by-light scattering amplitude to an
Axial-Vector-Vector triangle amplitude for which one has a good theoretical
understanding~\cite{MV04}. 

Unfortunately, the two asymptotic QCD constraints mentioned above are not
sufficient for a full model independent evaluation of the HLbL
contribution. Most of the last decade calculations found in the literature
are compatible with the QCD chiral and large-$N_c$ limits. They all
incorporate the $\pi^0$-exchange contribution modulated by $\pi^0 \gamma^*
\gamma^*$  form factors correctly normalized to the Adler, Bell-Jackiw
point-like coupling. 
They differ, however, on whether or not they satisfy the particular OPE
constraint mentioned above, and  
in the shape of the vertex form factors which follow from the different
models.

\begin{figure}[h!]
\begin{center}
\subfigure[]
{  \includegraphics[width=0.35\textwidth]{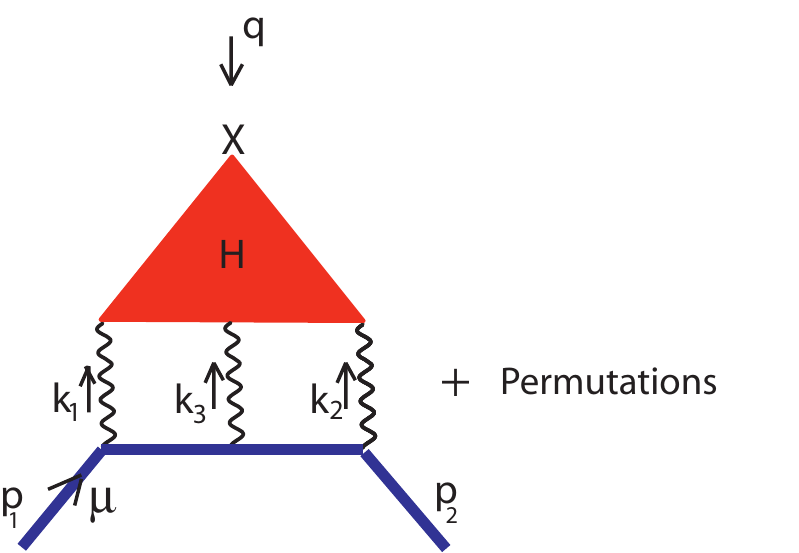}}
\subfigure[]
{  \includegraphics[width=0.35\textwidth]{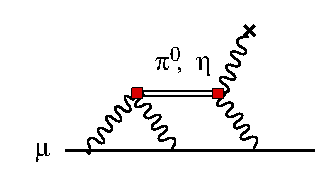}}
\end{center}
  \caption{(a)The Hadronic Light-by contribution. (b) The pseudoscalar meson contribution. }
\label{fg:HLBL}
\end{figure}

A synthesis of the model contributions, which was agreed to by
authors from each of the leading groups that have been working in this field, 
can be found in ref.~\cite{Prades:2010}\footnote{This compilation is
  generally referred 
  to as the ``Glasgow Consensus'' since it grew out of a workshop in 
Glasgow in 2007. http://www.ippp.dur.ac.uk/old/MuonMDM/ }. They
obtained
\be a_\mu^{\rm HLbL} = (105 \pm 26) \times 10^{-11} \, .
\label{eq:HLbL} \ee
An alternate evaluation~\cite{Jegerlehner:2009ry,Nyffeler:2009tw} 
obtained, $a_{\mu}^{\rm HLbL} =
(116\pm 40)\times 10^{-11}$, which agrees well with the Glasgow 
Consensus~\cite{Prades:2010}. 
Additional work on this contribution is underway on a number of fronts,
including on the lattice.   A workshop was held in March 2011 at the
Institute for Nuclear Theory in Seattle~\cite{INT:2011} which brought
together almost all of the interested experts.  This will be followed by a
workshop at the Mainz Institute for Theoretical Physics in April 2014.

One important point should be stressed here.  The main physics of the hadronic
light-by-light scattering contribution is well understood. In fact, but for
the sign error unraveled in 2002, the theoretical predictions for
$a_\mu^{\rm HLbL}$ have been relatively stable for more than ten
years\footnote{A calculation using a Dyson-Schwinger
  approach~\cite{Goecke:2011}  initially
  reported a much larger value for the HLbL contribution. Subsequently a
  numerical mistake was found. These authors are continuing this work, but
  the calculation is still incomplete.}.

\subsection{Summary of the Standard-Model Value and Comparison with
Experiment}


We determine the SM value using the new QED calculation from 
Aoyama~\cite{Aoyama1:2012}; the electroweak from Ref.~\cite{Miller:2012}, the
hadronic light-by-light contribution from the ``Glasgow 
Consensus''~\cite{Prades:2010}; and lowest-order hadronic contribution from
Davier, et al.,~\cite{Davier11}, or Hagiwara et al.,~\cite{Hagiwara:2011},
and the higher-order hadronic contribution from Ref.~\cite{Hagiwara:2011}.
   A summary of these values is given in
Table~\ref{tb:SMvalue}.

\begin{table}[h!]
\begin{center}
\caption{Summary of the Standard-Model contributions to the muon anomaly. Two
values are quoted because of the two recent evaluations of the lowest-order
hadronic vacuum polarization.}
{\small
\begin{tabular}{lr}
\hline \hline {\sc\small  } &
{\sc\small  Value $(\times \, 10^{-11})$ units  }
\\ \hline
QED ($\gamma + \ell$) & $116\,584\,718.951\pm 0.009\pm 0.019 \pm 0.007 \pm
0.077_{\alpha}$ \\
HVP(lo)~\cite{Davier11} &
$6\,923 \pm
42$\\
HVP(lo)~\cite{Hagiwara:2011} &
$6\,949 \pm
43$\\
HVP(ho)~\cite{Hagiwara:2011} &  $-98.4\pm 0.7 $ \\
HLbL&  $105\pm 26 $ \\
EW & $154\pm 1 $ \\
 \hline
 Total SM~\cite{Davier11} & $116\,591\,802 \pm 42_{\mbox{\rm \tiny H-LO}} 
\pm 26_{\mbox{\rm \tiny H-HO}}  \pm 2_{\mbox{\rm \tiny other}}   \, (\pm 49_{\mbox{\rm \tiny tot}}) $\\
  Total SM~\cite{Hagiwara:2011} & $116\,591\,828 \pm 43_{\mbox{\rm \tiny H-LO}} 
\pm 26_{\mbox{\rm \tiny H-HO}}  \pm 2_{\mbox{\rm \tiny other}}   \, (\pm 50_{\mbox{\rm \tiny tot}}) $\\
 \hline\hline\
\end{tabular}}
\label{tb:SMvalue}
\end{center}
\end{table}

This SM value is to be compared with the combined $a_\mu^+$ and
$a_\mu^-$ values from E821~\cite{Bennett:2006I} corrected for the
revised value of $\lambda= \mu_\mu/\mu_p$ from Ref~\cite{CODATA08I},
\be
a_\mu^{\rm E821} = (116\,592\,089 \pm 63)\times 10^{-11}\quad 
{\rm (0.54\,ppm)},\\
 \ee 
which give a difference of 
\bea 
\Delta a_{\mu}({\rm E821-SM}) &=& (287 \pm 80 )  \times 10^{-11} \ 
\cite{Davier11} \\
&=&  (261 \pm 78 )  \times 10^{-11} \ \cite{Hagiwara:2011}
\label{eq:Delta}
\eea
depending on which evaluation of the lowest-order hadronic contribution that
is used~\cite{Davier11,Hagiwara:2011}.

\begin{figure}[h!]
\begin{center}
\subfigure[]
{\includegraphics[width=0.4\textwidth]{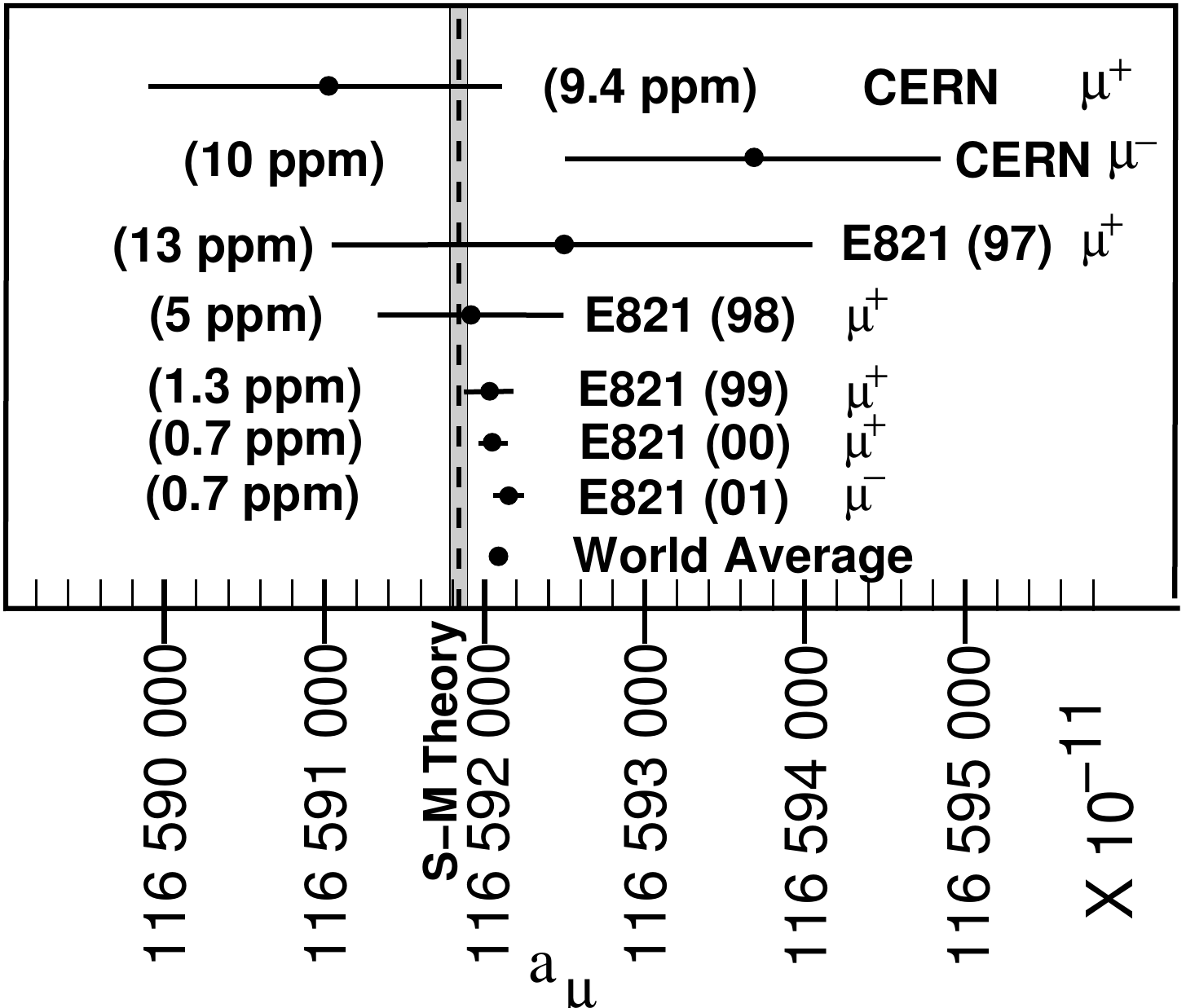}}
\subfigure[]
{\includegraphics[width=0.55\textwidth]{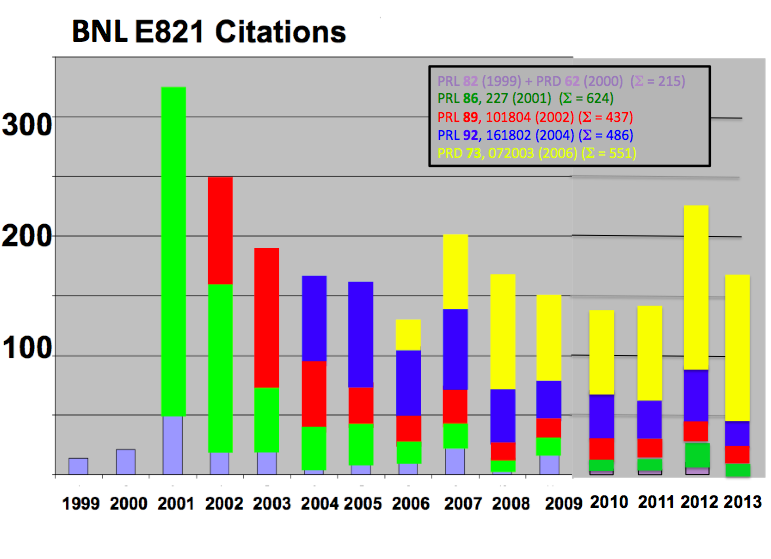}}
\end{center}
  \caption{(a)Measurements of $a_\mu$ from CERN and BNL E821.
The vertical band is the  SM value using the hadronic contribution from
Ref.~\cite{Davier11} (see Table~\ref{tb:SMvalue}). (b) Citations to the E821
papers by year. 
}
\label{fg:SM-Exp}
\end{figure}

 This comparison between the experimental values and the present
  Standard-Model value is shown
graphically in Fig. \ref{fg:SM-Exp}.
The lowest-order hadronic evaluation of Ref.~\cite{Benayoun:2012b}
using the hidden local symmetry model results in a
 difference between experiment and theory that ranges 
 between 4.1 to 4.7$\sigma$.

This difference of { 3.3} to 3.6 standard deviations is tantalizing, but we
emphasize that whatever the final agreement between the measured and
SM value turns out to be, it will have significant implications on
the interpretation of new phenomena that might be found at the LHC
and elsewhere. Because of the power of $a_\mu$ to constrain, or point to,
speculative models of New Physics, the E821 results have been highly cited,
with more than 2450 citations to date.

\section{Expected Improvements in the Standard-Model Value}

The present uncertainty on the theoretical value is dominated by the hadronic
contributions~\cite{Davier11,Hagiwara:2011} (see Table~\ref{tb:SMvalue}). 
The lowest-order contribution  determined from $e^+e^-
\rightarrow {\rm hadrons}$ data using a dispersion relation is theoretically
relatively straightforward.
 It does require the combination of data sets from different
experiments. 
The only significant theoretical uncertainty comes from radiative
corrections, such as vacuum polarization (running $\alpha$), along with
 initial and
final state radiation effects, which are needed to obtain the correct hadronic
cross section at the required level of precision. 
This was a problem for the older data sets. In the analysis of the
data collected over the past 15 years, which now dominate the determination
of the hadronic contribution, the
treatment of radiative corrections has been significantly
 improved. Nevertheless, an additional uncertainty due to the treatment of these
radiative corrections in the older data sets has been estimated to be of
the order of $20\times 10^{-11}$~\cite{Hagiwara:2011}.  As more data
become available, this uncertainty will be significantly reduced. 

There are two methods that have
been used to measure the hadronic cross sections: 
The energy scan (see Fig.~\ref{fg:hadpro}(b)), and 
using initial state radiation with a fixed beam
energy to measure the  cross section for  energies below the total
center-of-mass energy of the colliding beams (see Fig.~\ref{fg:hadpro}(c)). 
Both are being employed in the next round of measurements.
The data from the new experiments that are now underway at VEPP-2000 in
Novosibirsk
and  BESIII in Beijing, when combined with the analysis of
existing multi-hadron final-state data from BaBar and Belle,
should significantly reduce the uncertainty on the
lowest-order hadronic contribution.

The hadronic-light-by-light contribution does not lend itself to
determination by a dispersion relation. Nevertheless there are some
experimental data that can help to pin down related amplitudes and to
constrain form factors used in the model calculations. 

\subsection{Lowest-order Hadronic Contribution}

Much experimental and theoretical work is going on worldwide to
refine the hadronic contribution. The theory of $(g-2)$, relevant 
experiments to determine the hadronic contribution, including work on the
lattice,  have featured prominently
in the series of tau-lepton workshops and PHIPSI workshops which are
held in alternate years.
Over the development period of Fermilab E989, we expect further
improvements in the SM-theory evaluation.  This projection is based
on the following developments:

\subsubsection {Novosibirsk}
The VEPP2M machine has been upgraded to VEPP-2000. The maximum
energy has been increased from $\sqrt{s} =1.4$ GeV to 2.0 GeV.
Additionally, the SND detector has been upgraded and the CMD2
detector was replaced by the much-improved CMD3 detector. The
cross section will be measured from threshold to 2.0~GeV using an
energy scan, filling in the energy region between 1.4~GeV, where the
previous scan ended, up to 2.0~GeV, the lowest energy point reached
by the BES collaboration in their measurements.  See
Fig.~\ref{fg:had-cont} for the present contribution to the overall
error from this region. Engineering runs began in 2009, and data
collection started in 2011. So far two independent energy scans between
1.0 and 2.0~GeV were performed in 2011 and 2012. The peak luminosity
of $3\times 10^{31} {\rm cm}^{-2} {\rm s}^{-1}$ 
was achieved, which was limited by the positron 
production rate. The new injection facility, scheduled
 to be commissioned during
the 2013-2014 upgrade, should permit the
luminosity to reach $10^{32} {\rm cm}^{-2} {\rm s}^{-1}$ .
 Data collection  resumed
in late 2012 with a new energy scan over
energies below 1.0~GeV. The goal of experiments at VEPP-2000 is to
achieve a systematic error 0.3-0.5\% in the $\pi^+\pi^-$ channel,
 with negligible
statistical error in the integral. The high statistics, expected at VEPP-2000,
should allow a detailed comparison of the measured cross-sections
with ISR results at BaBar and DA$\Phi$NE. After the upgrade, experiments at
VEPP-2000 plan to take a large amount
 of data at 1.8-2~GeV, around the $N\bar{N}$ threshold.
This will permit ISR data with the beam energy of 2~GeV, which is
 between the PEP2 energy at
the $\Upsilon (4S)$ and the 1~GeV $\phi$ energy at the DA$\Phi$NE
facility in Frascati. The dual ISR and scan approach will provide an
important cross check on the two central methods used to determine
the HVP.

\subsubsection{The BESIII Experiment}
The BESIII experiment at the Beijing tau-charm
factory BEPC-II has already collected several femtobarns of integrated
luminosity 
at various centre-of-mass energies in the range 3 - 4.5 GeV.
The ISR program includes cross section measurements of:
$e^+e^- \to \pi^+\pi^-$, $e^+e^- \to \pi^+\pi^-\pi^0$, 
$e^+e^- \to \pi^+\pi^-\pi^0\pi^0$ --
the final states most relevant to $(g-2)_\mu$. 
Presently, a data sample of 2.9 fb$^{-1}$ 
at $\sqrt{s}=3.77$ GeV is being analyzed, but new data at $\sqrt{s} > 4$ GeV can be used for 
ISR physics as well and will double the statistics.
Using these data, hadronic invariant masses from threshold up to 
approximately 3.5 GeV can be accessed at BESIII.
Although the integrated luminosities 
are orders of magnitude lower compared to the $B$-factory
experiments BaBar and BELLE, the ISR method at BESIII still 
provides competitive statistics. This is due to the fact that the
most interesting mass range for the HVP contribution of $(g-2)_\mu$, which is
below approximately 3 GeV, 
is very close to the centre-of-mass energy of the collider BEPC-II and hence
leads 
to a configuration where
only relatively low-energetic ISR photons need to be emitted, providing
a high ISR cross section.
Furthermore, in contrast to the $B$ factories, small angle ISR photons can be 
included in the event selection for kinematic reasons which leads to a very
 high overall geometrical acceptance. 
Compared to the KLOE experiment, background from final state radiation (FSR)
is reduced significantly as 
this background decreases with increasing center of mass energies of the collider.  
BESIII is aiming for a precision measurement of the ISR $R$-ratio 
$R_{\rm ISR}= N(\pi\pi\gamma)/N(\mu\mu\gamma)$ with a precision of
about 1\%. This requires an excellent pion-muon separation, which
is achieved by training a multi-variate neural network. 
As a preliminary result, an absolute cross section
measurement of the reaction $e^+e^- \rightarrow \mu^+\mu^- \gamma$
 has been achieved, which
agrees with the QED prediction within 1\% precision.
%

Moreover, at BESIII a new energy scan campaign is planned 
to measure the inclusive $R$ ratio in the energy range between 2.0 and 4.6 GeV.  Thanks to
the good performance of the BEPC-II accelerator and the BESIII detector 
a significant improvement upon the existing BESII measurement can be expected. 
The goal is to arrive at an inclusive 
$R$ ratio measurement with about 1\% statistical and 3\% systematic precision 
per scan point. 

\subsubsection{Summary of the Lowest-Order Improvements from Data}

A substantial amount of new $e^+e^-$ cross section
 data will become available over the next
few years.  These data have the potential to significantly reduce the 
error on the lowest-order hadronic contribution. 
These improvements
can be obtained by reducing the uncertainties of the hadronic
cross-sections from 0.7\% to 0.4\% in the region below 1 GeV and from 6\%
to 2\% in the region between 1 and 2 GeV as shown in
 Table~\ref{tb:LO-impr}.

\begin{table}[h]
\begin{center}
 \renewcommand{\arraystretch}{1.4}
 \setlength{\tabcolsep}{1.6mm}
{\footnotesize
\begin{tabular}{|c|c|c|c|c|}
\hline
 & $\delta (\sigma)/\sigma$ present &$\delta a_{\mu}$present
 & $\delta (\sigma)/\sigma$ future &$\delta a_{\mu}$future  \\
\hline
$\sqrt{s}<1$~GeV & 0.7\% & 33 & 0.4\% & 19 \\
$1<\sqrt{s}<2$~GeV & 6\% & 39 & 2\% & 13 \\
$\sqrt{s}>2$~GeV & & 12 & & 12 \\
\hline
total & & 53 & & 26 \\
\hline
   
\end{tabular}
}
\caption{\label{tb:LO-impr} Overall uncertainty of the cross-section
measurement required to get the reduction of uncertainty on $a_{\mu}$ in units
$10^{-11}$ for
three regions of $\sqrt{s}$ (from Ref.~\cite{Jegerlehner:2008zz}).}
\end{center}
\end{table}
\subsubsection {Lattice calculation of the Lowest-Order HVP:}
With computer power presently available, it is possible for lattice QCD calculations to make important
contributions to our knowledge of the lowest-order hadronic
contribution. Using several different discretizations for QCD, lattice groups
around the world are computing the
HVP~\cite{Burger:2013jya,DellaMorte:2012cf,Blum:2013qu,Boyle:2011hu,Aubin:2006xv}
(see also several recent talks at Lattice 2013 (Mainz). The varied
techniques have different systematic errors, but in the continuum limit
$a\to0$ they should all agree. Many independent calculations provide a
powerful check on the lattice results, and ultimately the dispersive ones
too.  

Several groups are now performing simulations with physical light quark
masses on large boxes, eliminating significant systematic errors. So called
quark-disconnected diagrams are also being calculated, and several recent
theory advances will help to reduce systematic errors associated with fitting
and the small $q^{2}$
regime~\cite{deDivitiis:2012vs,DellaMorte:2012cf,Blum:2012uh,Aubin:2012me,Feng:2013xsa,Aubin:2013daa}. While
the HVP systematic errors are well understood, significant computational 
resources are needed to control them at the $\sim1\%$ level, or
better. Taking into account current resources and those expected in the next
few years, the lattice-QCD uncertainty on $a_\mu$(HVP), currently at the
$\sim 5$\%-level, can be reduced to 1 or 2\% 
within the next few years. This is already 
interesting as a wholly independent check of the dispersive
results for $a_\mu$(HVP). With increasing experience and computer power, it
should 
be possible to compete with the $e^+e^-$ determination of
$a_\mu$(HVP) by the end of the decade, perhaps sooner with additional
technical advances. 

\subsection{The Hadronic Light-by-Light contribution}


There are two approaches to improving the HLbL contribution: Measurements of
$\gamma^*$ physics at BESIII and KLOE; Calculations on the lattice.
In addition to the theoretical work on the HLbL, 
the KLOE detector at DA$\Phi$NE has been upgraded with a tagging system to
detect the final leptons in the reaction $e^+ e^- \to e^+ e^-
\gamma^*\gamma^*$.
Thus a coincidence between the scattered electrons and a $\pi^0$
would provide information 
on $\gamma^* \gamma^* \rightarrow \pi^0$~\cite{KLOE-2}, 
and will provide experimental constraints on the
models used to calculate the hadronic light-by-light 
contribution~\cite{Babusci:2011bg}.

Any experimental information on the neutral pion lifetime and the
transition form factor is important in order to constrain the models
used for calculating the pion-exchange contribution (see
Fig.~\ref{fg:HLBL}(b)).   However, having
a good description, e.g.\ for the transition form factor, is only
necessary, not sufficient, in order to uniquely determine
$a_\mu^{\mathrm{HLbL};\pi^0}$. As stressed in
Ref.~\cite{Jegerlehner_off-shell}, what enters in the calculation of
$a_\mu^{{\rm HLbL}; \pi^0}$ is the fully off-shell form factor ${\cal
  F}_{{\pi^0}^*\gamma^*\gamma^*}((q_1 + q_2)^2, q_1^2, q_2^2)$ (vertex
function), where also the pion is off-shell with 4-momentum $(q_1 +
q_2)$. Such a (model dependent) form factor can for instance be
defined via the QCD Green's function $\langle VVP \rangle$, see
Ref.~\cite{Nyffeler:2009tw} for details.  The form factor with on-shell
pions is then given by ${\cal F}_{\pi^0\gamma^*\gamma^*}(q_1^2, q_2^2)
\equiv {\cal F}_{{\pi^0}^*\gamma^*\gamma^*}(m_\pi^2, q_1^2, q_2^2)$.
Measurements of the transition form factor ${\cal
  F}_{\pi^0\gamma^\ast\gamma}(Q^2) \equiv {\cal
  F}_{{\pi^0}^\ast\gamma^\ast\gamma^\ast}(m_{\pi}^2, -Q^2, 0)$ are in
general only sensitive to a subset of the model parameters and do not
permit the reconstruction the full off-shell form factor.

For different models, the effects of the off-shell pion can vary a
lot. In Ref.~\cite{Nyffeler:2009tw} the off-shell lowest meson dominance
 (LMD) plus vector meson dominance (LMD+V) form factor was
proposed and the estimate $a_{\mu; {\rm LMD+V}}^{{\rm HLbL}; \pi^0} =
(72 \pm 12) \times 10^{-11}$ was obtained (see also Ref.~\cite{KN_EPJC_2001}). The error estimate comes
from the variation of all model parameters, where the uncertainty of
the parameters related to the off-shellness of the pion completely
dominates the total error. In contrast to the off-shell LMD+V model,
many other models, e.g.\ the VMD model or constituent quark models, do
not have these additional sources of uncertainty related to the
off-shellness of the pion. These models often have only very few
parameters, which can all be fixed by measurements of the transition
form factor or from other observables. Therefore, for such models, the
precision of the KLOE-2 measurement can dominate the total precision of
$a_\mu^{\mathrm{HLbL};\pi^0}$.

Essentially all evaluations of the pion-exchange contribution use for
the normalization of the form factor, ${\cal
  F}_{{\pi^0}^*\gamma^*\gamma^*}(m_\pi^2, 0, 0) = 1 / (4 \pi^2
F_\pi)$, as derived from the Wess-Zumino-Witten (WZW) term. Then the
value $F_\pi = 92.4~\mev$ is used without any error attached to it,
i.e. a value close to $F_\pi = (92.2 \pm 0.14)~\mbox{MeV}$, obtained
from $\pi^+ \to \mu^+ \nu_\mu(\gamma)$~\cite{Nakamura:2010zzi}. If one
uses the decay width $\Gamma_{\pi^0 \to \gamma\gamma}$ for the
normalization of the form factor, an additional source of uncertainty
enters, which has not been taken into account in most
evaluations~\cite{Nyffeler:2009uw}. 
Until recently, the experimental world average of 
$\Gamma^{PDG}_{\pi^0 \to \gamma\gamma}= 7.74\pm 0.48$ eV~\cite{Nakamura:2010zzi} was only known
to 6.2\% precision. 
Due to the poor agreement between the existing data, the PDG error of the
width average is inflated (scale factor of $2.6$) and it gives an additional motivation for new precise
measurements. The PrimEx Collaboration, using a Primakoff effect
 experiment at JLab, has achieved 2.8\% fractional 
precision~\cite{Larin:2010kq}.
There are plans to further reduce the
uncertainty to the percent level.
Though theory and experiment are in a fair agreement, a better experimental precision is needed to really test the theory predictions.

\subsubsection{Impact of KLOE-2 measurements on $a_\mu^{\mathrm{HLbL};\pi^0}$}
For the new data taking of the KLOE-2 detector,
 which is expected to start by the end of 2013,
new small angle
 tagging detectors have been installed along DA$\Phi$NE beam line.

These ``High Energy Tagger'' detectors \cite{het} 
offer the possibility to study 
a program of $\gamma\gamma$ physics through the process $e^+e^-\to 
e^+\gamma^* e^-\gamma^* \to e^+e^- X$. 

\begin{figure}[h]
\centerline{\includegraphics[width=.7\textwidth]{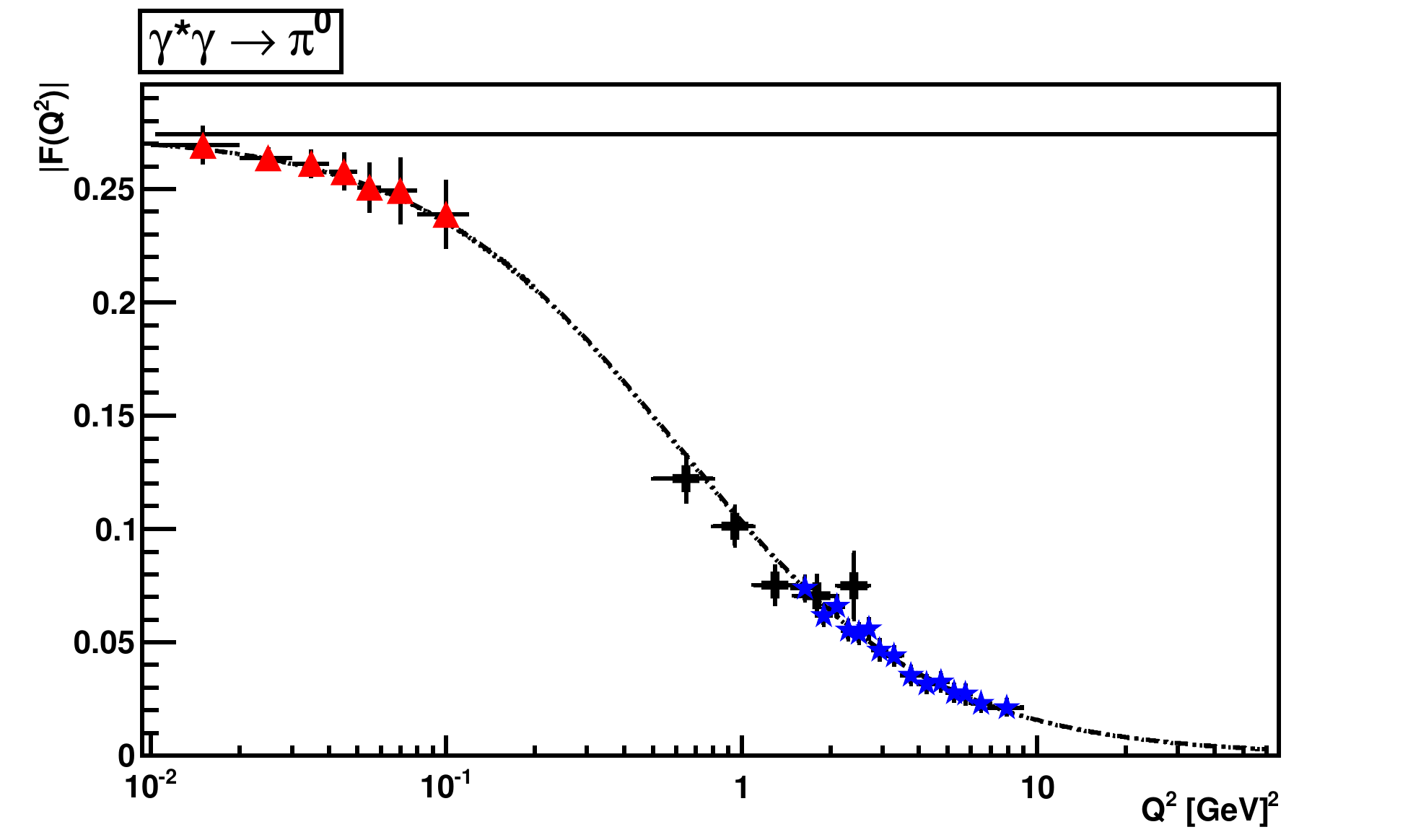}}
\caption{Simulation of KLOE-2 measurement of $F(Q^2)$ (red triangles)
  with statistical errors for $5$~fb$^{-1}$, corresponding to one year
  of data taking. The dashed line is the $F(Q^2)$ form factor
  according to the LMD+V model~\cite{Nyffeler:2009tw,KN_EPJC_2001}, the solid
  line is $F(0) = 1/(4\pi^2  F_\pi)$ given by the Wess-Zumino-Witten term.
 Data~\cite{TFF_data}
  from CELLO (black crosses) and CLEO (blue stars) at high $Q^2$ are
  also shown for illustration.}
\label{simulation_FF_data}
\end{figure}

In Ref.~\cite{KLOE-2_impact} it was shown that planned measurements at
KLOE-2 could determine the $\pi^0\to\gamma\gamma$ decay width to 1\%
statistical precision and the $\gamma^\ast\gamma\to\pi^0$ transition
form factor ${\cal F}_{\pi^0\gamma^\ast\gamma}(Q^2)$ for small
space-like momenta, $0.01~\mbox{GeV}^2 \leq Q^2 \leq
0.1~\mbox{GeV}^2$, to 6\% statistical precision in each bin. The
simulations have been performed with the Monte-Carlo program
EKHARA~\cite{EKHARA} for the process $e^+ e^- \to e^+ e^- \gamma^*
\gamma^* \to e^+ e^- \pi^0$, followed by the decay $\pi^0 \to
\gamma\gamma$ and combined with a detailed detector simulation. The
results of the simulations are shown in
Figure~\ref{simulation_FF_data}. The KLOE-2 measurements will allow to
almost directly measure the slope of the form factor at the origin and
check the consistency of models which have been used to extrapolate
the data from larger values of $Q^2$ down to the origin.
With the decay width $\Gamma_{\pi^0\to\gamma\gamma}^{\rm PDG}$
$[\Gamma_{\pi^0\to\gamma\gamma}^{\rm PrimEx}]$ and current data for
the transition form factor ${\cal F}_{\pi^0\gamma^\ast\gamma}(Q^2)$,
the error on $a_\mu^{{\rm HLbL}; \pi^0}$ is $\pm 4 \times 10^{-11}$
[$\pm 2 \times 10^{-11}$], not taking into account the uncertainty
related to the off-shellness of the pion. Including the simulated
KLOE-2 data reduces the error to $\pm (0.7 - 1.1) \times
10^{-11}$. 

\subsubsection{BESIII Hadronic light-by-light contribution} 

Presently, data taken at $\sqrt{s}=$3.77 GeV are being analyzed 
to measure the form factors of the 
reactions $\gamma^*\gamma \to X$, where
$X=\pi^0, \eta, \eta^\prime, 2\pi$.

BESIII has launched a program of two-photon interactions with
the primary goal to measure the transition form factors (TFF)
of pseudoscalar mesons as well as of the two-pion system in the
spacelike domain. These measurements are carried out in the
single-tag mode, i.e. by tagging one of the two beam leptons at large
polar angles
and by requiring that the second lepton is scattered at small
polar angles. With these kinematics
 the form factor, which in general depends on the
virtualities of the two photons, reduces to $F(Q^2)$,
where $Q^2$ is the negative momentum transfer of
the tagged lepton. At BESIII, the process $\gamma \gamma^* \to \pi^0$, which
is known to play 
a leading contribution in the HLbL correction to $(g-2)$, can be measured
with unprecedented 
precision in the 
$Q^2$ range between 0.3 GeV$^2$ and 4 GeV$^2$. In the future
BESIII will also embark on untagged as well as double-tag measurements, in
which either  
both photons are quasi-real or feature a high virtuality. The goal is to
carry out 
this program for the final states $\pi^0, \eta, \eta^\prime, \pi\pi$.
It still needs to be proven that the 
small angle detector,
which recently has been installed close to the BESIII beamline, 
can be used for the two-photon program.
\subsubsection {Lattice calculation of Hadronic Light-by-Light Scattering:}
Model calculations show that the hadronic light-by-light (HLbL) contribution
is roughly $(105 
\pm 26)\times 10^{-11}$, $\sim 1$ ppm of $a_{\mu}$. Since the error
attributed to this estimate is difficult to reduce, a modest, but
first principles  
calculation on the lattice would have a large impact.
Recent progress towards this goal has been reported~\cite{Blum:2013qu}, where
a non-zero signal (statistically speaking) for a part of the amplitude
emerged in the same ball-park as the model estimate. The result was computed
at non-physical quark mass, with other systematic errors mostly
uncontrolled. Work on this method, which treats both QED and QCD interactions
non-perturbatively, is continuing. The next step is to repeat the calculation
on an ensemble of gauge configurations that has been generated with
electrically charged sea quarks (see the poster by Blum presented at Lattice
2013). The charged sea quarks automatically include the quark disconnected
diagrams that were omitted in the original calculation and yield the complete
amplitude. As for the HVP, the computation of the HLbL contribution requires
significant resources which are becoming available. While only one group has
so far attempted the calculation, given the recent interest in the HVP
contribution computed in lattice QCD and electromagnetic corrections to
hadronic observables in general, it seems likely that others will soon enter
the game. And while the ultimate goal is to compute the HLbL contribution to
10\% accuracy, or better,  
we emphasize that a lattice calculation with even a solid 30\% error would
already be very interesting. Such a result, while not guaranteed, is not out
of the question during the next 3-5 years.  

\section{Summary}

The muon and electron anomalous magnetic moments are among, if not the most
precisely measured and calculated quantities in all of physics.  The
theoretical uncertainty on the Standard-Model contribution
to $a_\mu$ is $\simeq 0.4$~ppm,
slightly smaller than the experimental error from BNL821.  The new Fermilab
experiment, E989, will achieve a precision of 0.14~ppm.  While the hadronic
corrections will most likely not reach that level of precision, their
uncertainty will be significantly decreased.  The lowest-order contribution
will be improved by new data from Novosibirsk and BESIII. On the timescale of
the first results from E989, the lattice will also become relevant.

The hadronic light-by-light contribution will also see significant
improvement.  The measurements at Frascati and at BESIII will provide
valuable experimental input to constrain the model calculations.  There is
 hope that the lattice could produce a meaningful result by 2018.

\begin{figure}[h!]
\begin{center}
\begin{minipage}{\textwidth-.2in}
  \begin{minipage}{\textwidth/2}
{ 
\begin{tabular}{|c|c c |c|}
\hline
 Error &  \cite{Davier11} & \cite{Hagiwara:2011} & Future \\
\hline
 $\delta a_{\mu}^{\rm
SM}$ & 49 & 50 & 35 \\
\hline
 $\delta a_\mu^{\rm HLO}$ & 42 & 43 & 26 \\
$\delta a_\mu^{\rm HLbL}$  & 26 &26 &  25 \\
\hline
$\delta (a_\mu^{\rm EXP} - a_\mu^{\rm SM})$ & 80 & 80 & 40 \\
\hline   
\end{tabular}
}
  \end{minipage}
\hskip0.2in
  \begin{minipage}{\textwidth/2}
    \begin{center}
\includegraphics[width=.9\textwidth]{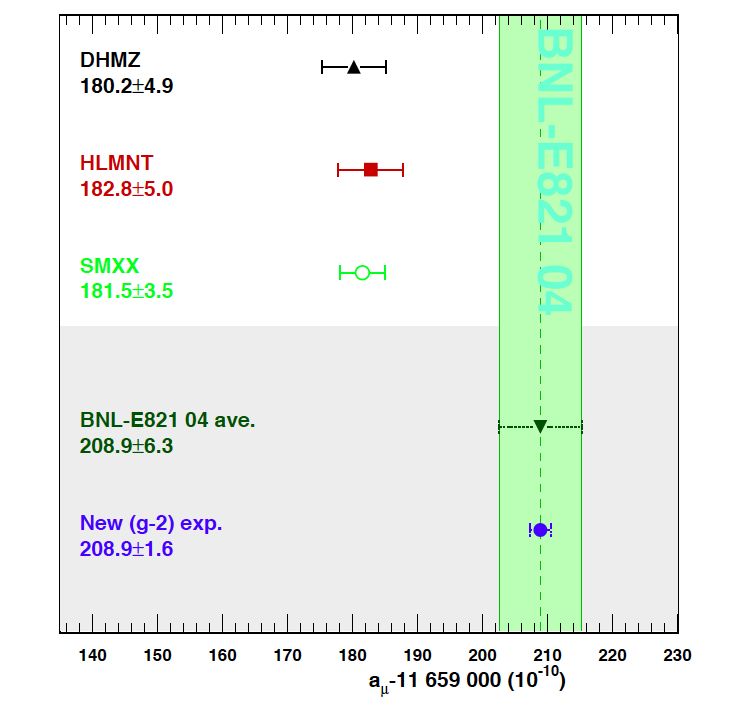}
    \end{center}
  \end{minipage}
 \end{minipage}
\end{center}
\caption{\label{tab:g-2a} Estimated uncertainties $\delta a_{\mu}$ in units
of $10^{-11}$ according to Refs.~\cite{Davier11,Hagiwara:2011} and
(last column) prospects for improved precision in the $e^+e^-$
hadronic cross-section measurements.
The final  row projects the uncertainty on
the difference with the Standard Model,  $\Delta a_{\mu}$. 
The figure give the comparison between $a_{\mu}^{\rm SM}$ and $a_{\mu}^{\rm EXP}$. 
DHMZ is Ref.~\cite{Davier11}, HLMNT is Ref.~\cite{Hagiwara:2011}; 
``SMXX'' is the same central value with a
reduced error as 
 expected by the improvement on the hadronic cross section measurement (see
 text); ``BNL-E821 04 ave.'' is the current experimental value 
of $a_{\mu}$;  ``New (g-2) exp.'' is the same central value with a fourfold
improved precision as planned by the future (g-2)
experiments at Fermilab and J-PARC.
}
\end{figure}

We summarize possible near-future improvements in the table below.
Since it is difficult to project the improvements in the hadronic
light-by-light contribution, we assume a conservative improvement: That 
 the large amount of work
 that is underway
 to understand this contribution, both
 experimentally and on the lattice, will support the level of
 uncertainty assigned in the ``Glasgow  Consensus''.
With these improvements, the overall uncertainty on $\Delta a_{\mu}$ could be
reduced by a factor 2. In case the central value would remain the same, the
statistical significance would become 7-8 standard deviations, as it can be
seen in Fig.~\ref{tab:g-2a}.

Thus the prognosis is excellent that the results from E989 will clarify
whether the measured value of $a_\mu$ contains contributions from outside of
the Standard Model.  Even if there
 is no improvement on the hadronic error, but
the central theory and experimental values remain the same, the significance
of the difference would be over $5\sigma$.  However, with the worldwide
effort to improve on the Standard-Model value, it is most likely that the
comparison will be even more convincing.


\begin{thebibliography}{90}

\bibitem{Schwinger:1948} 
J. Schwinger, Phys. Rev. {\bf 73} (1948) 416, and Phys. Rev. {\bf 76}
(1949) 790. The former paper contains a misprint in the expression for
$a_e$ that is corrected in the longer paper. 

\bibitem{Aoyama1:2012} 
T. Aoyama, M. Hayakawa, T. Kinoshita and M. Nio,
Phys. Rev. Lett. {\bf 109} (2012) 111808.

\bibitem{Miller:2012} 
J. P. Miller, E. de Rafael, B. L. Roberts and D. St\"ockinger, 
Ann. Rev. Nucl. Part. Sci. {\bf 62} (2012) 237.

\bibitem{Stoeckinger2010} 
D. St\"ockinger, 
in Advanced Series on Directions in High Energy Physics - Vol. 20 
\textit{Lepton Dipole Moments}, eds.\ B. L. Roberts and W. J. Marciano, 
World Scientific (2010), p.393.

\bibitem{Hanneke08} 
D. Hanneke, S. Fogwell and G. Gabrielse, 
Phys. Rev. Lett. {\bf 100} (2008) 120801.

\bibitem{Bennett:2006I} 
G. W. Bennett et al. (The $g-2$ Collab.), Phys. Rev. {\bf D73} (2006) 072003.

\bibitem{Davier-LM} 
M. Davier, 
in Advanced Series on Directions in High Energy Physics - Vol. 20 
\textit{Lepton Dipole Moments}, eds.\ B. L. Roberts and W. J. Marciano, 
World Scientific (2010), chapter 8.

\bibitem{Bouchendira:2011} 
R. Bouchendira, P. Clade, S. Guellati-Khelifa, F. Nez and F. Biraben, 
Phys. Rev. Lett. {\bf 106} (2011) 080801.

\bibitem{CKM96} 
A. Czarnecki, B. Krause and W. J. Marciano, 
Phys. Rev. Lett. {\bf 76} (1996) 3267.

\bibitem{PPdR95} 
S. Peris, M. Perrottet and E. de Rafael, 
Phys. Lett. {\bf B355} (1995) 523.

\bibitem{CKM95} 
A. Czarnecki, B. Krause and W. Marciano, 
Phys. Rev. {\bf D52} (1995) 2619.

\bibitem{CMV03} 
A. Czarnecki, W. J. Marciano and A. Vainshtein,
Phys. Rev. {\bf D67} (2003) 073006, Erratum-ibid. {\bf D73} (2006) 119901. 

\bibitem{CM-LM} 
A. Czarnecki and W. J. Marciano,
in Advanced Series on Directions in High Energy Physics - Vol. 20 
\textit{Lepton Dipole Moments}, eds.\ B. L. Roberts and W. J. Marciano, 
World Scientific (2010), p. 11, and references therein.

\bibitem{Gnendiger:2013} 
C. Gnendiger, D. St\"ockinger and H. St\"ockinger-Kim,
Phys. Rev. {\bf D88} (2013) 053005.

\bibitem{weak} 
W. A. Bardeen, R. Gastmans and B. Lautrup, 
Nucl. Phys. {\bf B46} (1972) 319; 
R. Jackiw and S. Weinberg, 
Phys. Rev. {\bf D5} (1972) 157; 
I. Bars and M. Yoshimura, 
Phys. Rev. {\bf D6} (1972) 374;
K. Fujikawa, B. W. Lee and A. I. Sanda, 
Phys. Rev. {\bf D6} (1972) 2923.

\bibitem{Miller:07I} 
J. P. Miller, E. de Rafael and B. L. Roberts,
Rept. Prog. Phys. {\bf 70} (2007) 795.

\bibitem{KPPdeR02} 
M. Knecht, S. Peris, M. Perrottet and E. de Rafael,
JHEP {\bf 0211} (2002) 003.

\bibitem{Vain03} 
A. Vainshtein, Phys. Lett. {\bf B569} (2003) 187.
                                
\bibitem{KPPdeR04} 
M. Knecht, S. Peris, M. Perrottet and E. de Rafael, 
JHEP {\bf 0403} (2004)  035.

\bibitem{Davier11} 
M. Davier, A. Hoecker, B. Malaescu and Z. Zhang,
Eur. Phys. J.  {\bf C71} (2011) 1515, Erratum-ibid. {\bf C72} (2012) 1874.

\bibitem{Hagiwara:2011} 
K. Hagiwara, R. Liao, A. D. Martin, D. Nomura and T. Teubner, 
J. Phys. {\bf G38} (2011) 085003.

\bibitem{HMNT07} 
K. Hagiwara, A. D. Martin, D. Nomura and T. Teubner,
 Phys. Lett. {\bf B649} (2007) 173.

\bibitem{Davier07} 
M. Davier,
Nucl. Phys. Proc. Suppl. {\bf 169} (2007) 288.

\bibitem{Jegerlehner:2009ry} 
F. Jegerlehner and A. Nyffeler, Phys.Rept. {\bf 477} (2009) 1.

\bibitem{Alemany:1998} 
R. Alemany, M. Davier and A. H\"ocker, 
Eur. Phys. J. {\bf C2} (1998) 123.

\bibitem{Jegerlehner:2011} 
F. Jegerlehner and R. Szafron, 
Eur. Phys. J. {\bf C71} (2011) 1632.

\bibitem{Benayoun:2012a} 
M. Benayoun, P. David, L. DelBuono and F. Jegerlehner, 
Eur. Phys. J. {\bf C72} (2012) 1848.

\bibitem{Benayoun:2012b} 
M. Benayoun, P. David, L. DelBuono and F. Jegerlehner, 
Eur. Phys. J. {\bf C73} (2013) 2453.

\bibitem{KNPdeR02} 
M. Knecht, A. Nyffeler, M. Perrottet and E. de Rafael, 
Phys. Rev. Lett. {\bf 88} (2002) 071802.
         
\bibitem{MV04} 
K. Melnikov and A. Vainshtein, 
Phys. Rev. {\bf D70} (2004) 113006.

\bibitem{Prades:2010} 
J. Prades, E. de Rafael and A. Vainshtein,
in Advanced Series on Directions in High Energy Physics - Vol. 20 
\textit{Lepton Dipole Moments}, eds.\ B. L. Roberts and W. J. Marciano, 
World Scientific (2010), p. 303; and arXiv:0901.0306v1 [hep-ph].

\bibitem{Nyffeler:2009tw} 
A. Nyffeler, Phys. Rev. {\bf D79} (2009) 073012.

\bibitem{INT:2011} 
http://www.int.washington.edu/PROGRAMS/11-47w/

\bibitem{Goecke:2011} 
T. Goecke, C. S. Fischer and R. Williams, 
Phys. Rev. {\bf D83} (2011) 094006, Erratum-ibid. {\bf D86} (2012)
099901; Phys. Rev. {\bf D87} (2013) 034013.

\bibitem{CODATA08I} 
P. J. Mohr, B. N. Taylor and D. B. Newell (CODATA recommended values),
Rev. Mod. Phys. {\bf 80} (2008) 633.


\bibitem{Eidelman:1995} 
S. Eidelman and F. Jegerlehner, 
Z. Phys. {\bf C67} (1995) 585.

\bibitem{Akhmetshin:2002} 
R. R. Akhmetshin et al. (CMD2 Collaboration), 
Phys. Lett. {\bf B527} (2002) 161.

\bibitem{Davier:2003} 
M. Davier, S. Eidelman, A. H\"ocker and Z. Zhang, 
Eur. Phys. J. {\bf C27} (2003) 497.

\bibitem{Hagiwara:2004} 
K. Hagiwara, A. D. Martin, D. Nomura and T. Teubner, 
Phys. Rev. {\bf D69} (2004) 093003.

\bibitem{Jegerlehner:2008zz} 
F. Jegerlehner, Nucl. Phys. Proc. Suppl. {\bf 181-182} (2008) 26.

\bibitem{Burger:2013jya} 
F. Burger, X. Feng, G. Hotzel, K. Jansen, M. Petschlies and D. B. Renner, 
arXiv:1308.4327 [hep-lat].

\bibitem{DellaMorte:2012cf} 
M. Della Morte, B. Jager, A. Juttner and H. Wittig, 
PoS LATTICE {\bf 2012} (2012) 175 [arXiv:1211.1159 [hep-lat]].

\bibitem{Blum:2013qu} 
T. Blum, M. Hayakawa and T. Izubuchi, 
PoS LATTICE {\bf 2012} (2012) 022 [arXiv:1301.2607 [hep-lat]].

\bibitem{Boyle:2011hu} 
P. Boyle, L. Del Debbio, E. Kerrane and J. Zanotti, 
Phys. Rev. {\bf D85} (2012) 074504.

\bibitem{Aubin:2006xv} 
C. Aubin and T. Blum, 
Phys. Rev. {\bf D75} (2007) 114502.

\bibitem{deDivitiis:2012vs} 
G. M. de Divitiis, R. Petronzio and N. Tantalo, 
Phys. Lett. {\bf B718} (2012) 589.

\bibitem{Blum:2012uh} 
T. Blum, T. Izubuchi and E. Shintani, 
arXiv:1208.4349 [hep-lat].

\bibitem{Aubin:2012me} 
C. Aubin, T. Blum, M. Golterman and S. Peris, 
Phys. Rev. {\bf D86} (2012) 054509.

\bibitem{Feng:2013xsa} 
X. Feng, S. Hashimoto, G. Hotzel, K. Jansen, M. Petschlies and D. B. Renner, 
Phys. Rev. {\bf D88} (2013) 034505.

\bibitem{Aubin:2013daa} 
C. Aubin, T. Blum, M. Golterman and S. Peris,
arXiv:1307.4701 [hep-lat].

\bibitem{KLOE-2} 
G. Amelino-Camelia et al. (KLOE-2 Collaboration),
Eur. Phys. J. {\bf C68} (2010) 619.

\bibitem{Babusci:2011bg} 
D. Babusci, H. Czyz, F. Gonnella, S. Ivashyn, M. Mascolo, R. Messi,
D. Moricciani and A. Nyffeler et al., 
Eur. Phys. J. {\bf C72} (2012) 1917.

\bibitem{Jegerlehner_off-shell} 
F. Jegerlehner, Acta Phys.\ Polon. {\bf B38} (2007) 3021; 
\textit{The anomalous magnetic moment of the muon}, Springer (2008).

\bibitem{KN_EPJC_2001} 
M. Knecht and A. Nyffeler, 
Eur. Phys. J. {\bf C21} (2001) 659.

\bibitem{Nakamura:2010zzi} 
K. Nakamura et al., J. Phys. {\bf G37} (2010) 075021.

\bibitem{Nyffeler:2009uw} 
A. Nyffeler, PoS {\bf CD09} (2009) 080 [arXiv:0912.1441 [hep-ph]].

\bibitem{Larin:2010kq} 
I. Larin et al., Phys. Rev. Lett. {\bf 106} (2011) 162303.

\bibitem{het} 
F. Archilli et al., Nucl. Instrum. Meth. {\bf A617} (2010) 266.

\bibitem{KLOE-2_impact} 
D. Babusci et al., Eur. Phys. J. {\bf C72} (2012) 1917.

\bibitem{EKHARA} 
H. Czy\.z and S. Ivashyn, Comput. Phys. Commun. {\bf 182} (2011) 1338.

\bibitem{TFF_data} 
H. J. Behrend et al., Z. Phys. {\bf C49} (1991) 401; 
J. Gronberg et al., Phys. Rev. {\bf D57} (1998) 33.

\end{thebibliography}
\end{document}